\documentclass[12pt,a4paper]{article}

\makeatletter 
\def\set@curr@file#1{%
  \begingroup
    \escapechar\m@ne
    \xdef\@curr@file{\expandafter\string\csname #1\endcsname}%
  \endgroup
}
\def\quote@name#1{"\quote@@name#1\@gobble""}
\def\quote@@name#1"{#1\quote@@name}
\def\unquote@name#1{\quote@@name#1\@gobble"}
\makeatother

\usepackage{graphicx}
\usepackage{xcolor}
\usepackage[normalem]{ulem}
\usepackage{url}
\usepackage{siunitx}
\usepackage[T1]{fontenc}
\usepackage{amssymb}
\usepackage{ulem}
\usepackage{amsmath}
\usepackage{bm}
\usepackage{authblk}
\usepackage[left=2.25cm, right=2.25cm, top=2cm, bottom=2.5cm]{geometry}

\date{}

\title{\textbf{Electron localization in periodically strained graphene}}

\author{
Davide Giambastiani$^{1,2}$, 
Francesco Colangelo$^{3}$,
Alessandro Tredicucci$^{1,2}$,
Stefano Roddaro$^{1,2}$ and 
Alessandro Pitanti$^{2}$
}

\affil{$^1$ Dipartimento di Fisica "E. Fermi", Università di Pisa, Largo B. Pontecorvo 3, I-56127 Pisa, Italy}
\affil{$^2$ NEST, CNR - Istituto Nanoscienze and Scuola Normale Superiore, piazza San Silvestro 12, I-56127 Pisa, Italy}
\affil{$^3$ Institute for Quantum Electronics, ETH Z\"urich, CH-8093 Z\"urich, Switzerland}

\begin{document}

\maketitle

\begin{abstract}
Pseudo-magnetic field (PMF) in deformed graphene has been proposed as a promising and flexible method to quantum-confine electronic states and create gaps in the local density of states. Motivated by this perspective, we numerically analyze various different configurations leading to electronic localization and band flattening in periodically strained graphene. In particular, we highlight the existence of a fine structure in the pseudo-Landau levels confined in large-PMF regions, the emergence of states confined to PMF {\em nodes} as well as of snake-like orbits. In our paper, we further analyze the importance of the relative rotation and asymmetry of the strain lattice with respect to the atomic lattice and show how it can be used to modulate the PMF periodicity and to create localized orbits far from the strain points. Possible implementations and applications of the simulated structures are discussed.
\end{abstract}

\section{Introduction}
The impact of the mechanical deformation on the electronic band structure of graphene can be effectively described in terms of a fictitious magnetic field, which is routinely called pseudo-magnetic field (PMF)~\cite{guinea2010,vozmediano2010,amorim2016}. The PMF offers a non-standard and promising method to tune the band structure of graphene. It can be exploited not only to induce the formation of pseudo Landau levels (pLLs)~\cite{levy2010,lu2012,meng2013}, but also to trap electrons in pseudo-magnetic quantum dots~\cite{qi2013,settnes2016pseudomagnetic}. Exploiting the pseudo-spin dependent action of PMFs~\cite{settnes2016pseudomagnetic,carrillo2014,manesco2020}, electron waveguides~\cite{carrillo2016} and valley-filtering devices~\cite{settnes2016graphene,milovanovic2016,carrillo2016,hsu2020} can be created. 

A particularly attractive perspective opened by the periodic PMFs is the possibility to induce a flat band dispersion~\cite{mao2020,milovanovic2020}, where electrons have a quenched kinetic energy and are more prone to form correlated states driven by electron-electron interactions\cite{cao2018,sharpe2019,yankowitz2019,kim2017}. Recently, periodic strain profiles have been demonstrated in twisted multilayers~\cite{liu2018,shi2020}, providing the cleanest possible implementation of the periodic PMF concept so far. Alternative approaches include the use of non-planar substrates~\cite{tomori2011,jiang2017,nigge2019,banerjee2020} as well as micrometric polymeric or metallic~\cite{shioya2014,shioya2015,colangelo2018,gilbert2019} actuators which are also used to strain two-dimensional materials different from graphene~\cite{colangelo2019,azizimanesh2021}. The latter methods yield flexible strain profiles and periodicity schemes, at the expense of the more invasive action of the micrometric actuators, which can strongly impact the graphene underneath.

Here, motivated by these results, we numerically analyze the different mechanisms and factors leading to quantum confinement of electronic states in periodically strained graphene and to the emergence of flat band dispersion. We show how the relative orientation between the atomic and strain lattices can lead to markedly different PMF patterns and band profiles. In addition, we correlate the emergence of flat dispersions with the electrons localization in the strain lattice. This allows to identify: pLL multiplets in regions of large PMF, delocalized snake-like states propagating in between regions of different PMF signs and modes confined in the PMF nodes. Furthermore, based on the possibility of controlling the strain on the individual sites, we simulate strain lattices with reduced symmetry and discuss the appearance of edge states localized outside the PMF lobes, with possible applications for the creation of electronic waveguides, similarly to what expected in case of folded graphene~\cite{carrillo2016}. 

This paper is organized as follows: in Section~\ref{sec:methods}, we describe the theoretical model which we used in our numerical analysis; in Section~\ref{sec:symm}, we investigate how the relative rotation between the atomic lattice of graphene and the strain lattice can modulate the resulting PMF pattern; in Section~\ref{sec:asymm}, we model a periodic stress lattice with two uneven stress sites per supercell and calculate the resulting electronic structure of the system. Finally, the conclusions are drawn in Section~\ref{sec:conclusion}.

\section{Model for strain-induced confinement} \label{sec:methods}

\begin{figure}[]
    \centering
    \includegraphics[scale=0.45]{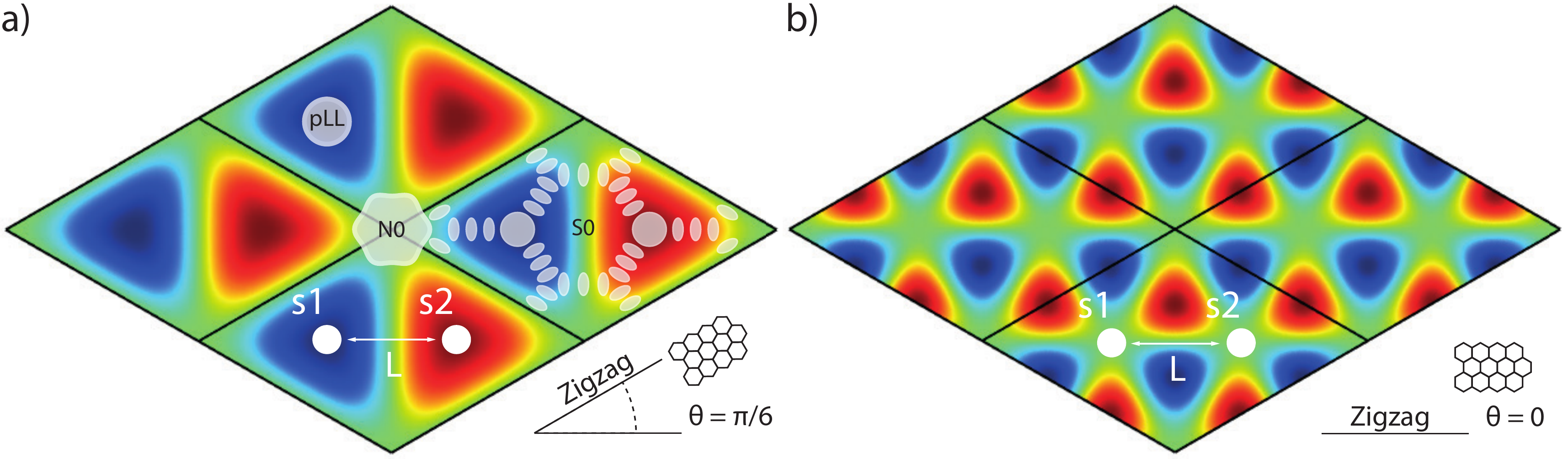}
    \caption{a) Primitive cell and pseudo-magnetic field $B_{ps}$ for $\theta=\pi/6$. The sites $s1$ and $s2$
    identify the two sublattices of the stress pattern. Schematic sketch of wavefunctions for N0, pLL and S0 states, as discussed in the main text. b) Same sketch of panel (a) for $\theta=0$.}
    \label{fig1}
\end{figure}

\begin{figure}[]
    \centering
    \includegraphics[scale=0.4]{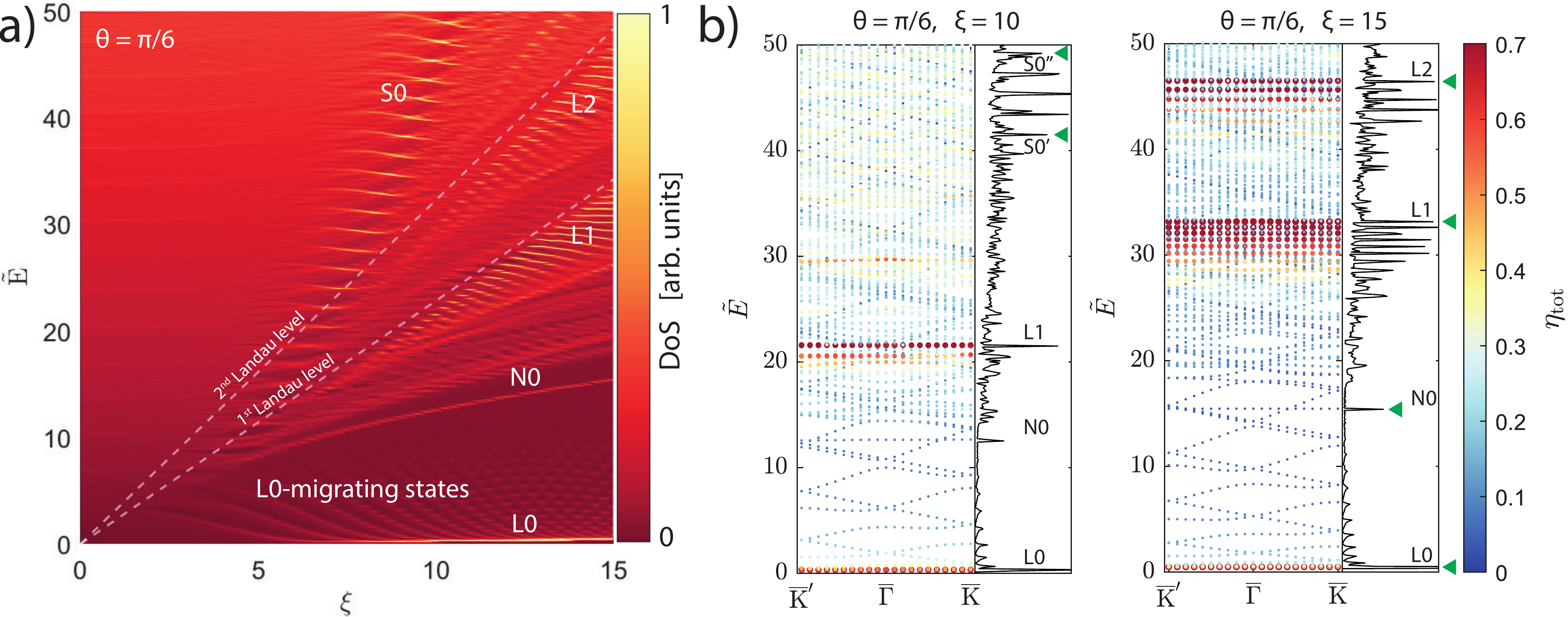}
    \caption{a) Normalized density of states as a function of adimensional energy $\tilde{E}$ and PMF ($\xi$). The white dashed lines are the $1^{st}$ and $2^{nd}$ Landau levels in graphene for real and uniform magnetic fields. b) Energy band structure and corresponding density of states (cross-section from panel (a)) for $\xi = 10$ (left) and $\xi = 15$ (right). The green triangles in the DoS correspond to the labeled states.}
    \label{fig2}
\end{figure}

The properties of electrons in deformed graphene near the Dirac points ($K$ and $K'$) can be described through the effective low-energy Hamiltonian~\cite{amorim2016,neto2009} 
\begin{equation}
\mathcal{H}=v_F
\begin{pmatrix}
\bm{\tau}\cdot(\mathbf{p}-e\mathbf{A}_{ps}) & 0 \\
0 & -\bm{\tau}\cdot(\mathbf{p}+e\mathbf{A}_{ps}),
\end{pmatrix}
\label{eq:dirac}
\end{equation}
for the 4-component spinor $\chi = (\psi_A^K,\psi_B^K,\psi_B^{K'},\psi_A^{K'})^T$, where $A$ and $B$ stand for the 2 pseudospin components, $v_F$ is the Fermi velocity, $e$ is the electronic charge, $\boldsymbol{\tau}=(\tau_x,\tau_y)$ is the vector of Pauli matrices and $\mathbf{p} = -i\hbar(\partial_x,\partial_y)^T$. The term $\mathbf{A}_{ps} = (A_x,A_y)$ can be seen as a vector potential producing a PMF $B_{ps}=\partial_x A_y-\partial_yA_x$. The potential $\mathbf{A}_{ps}$ is directly related to the mechanical deformation of graphene and can be expressed in terms of the strain tensor $\varepsilon$~\cite{guinea2010, verbiest2015} (see Supplementary Information):
\begin{equation}
\begin{split}
& A_x = \frac{\hbar \beta}{e a}[(\varepsilon_{xx}-\varepsilon_{yy})\cos(3\theta)-2\varepsilon_{xy}\sin(3\theta)] \\
& A_y = -\frac{\hbar \beta}{e a}[(\varepsilon_{xx}-\varepsilon_{yy})\sin(3\theta)+2\varepsilon_{xy}\cos(3\theta)],
\end{split}
\label{eq:pseudovector}
\end{equation}
where $\beta$ is the Gr\"{u}inesen parameter ($\beta \approx 2$)~\cite{settembrini2016}, $a$ is the lattice constant and $\theta$ is the angle between the x-axis and the zigzag direction in the graphene crystal (see figure~\ref{fig1}). \\
\indent In our numerical simulation, we generate a periodic strain pattern modeled using a set of Gaussian compressive stress profiles, which provides a first-order description of the effect of a circular actuator directly deposited on graphene (see Supplementary Information). Stress sites are arranged in an infinite honeycomb superlattice which allows exploring a large range of confinement scenarios depending on its orientation and asymmetry. Given this choice, the whole system can be described simulating two stress regions within the primitive supercell, which coincides with the simulation cell where we solve the eigenvalue equation $\mathcal{H}\chi=E\chi$, for the graphene valley $K$ and the pseudospin $A$ (see equation~\ref{eq:dirac}). \\
\indent The two nonequivalent sites ($s1$ and $s2$) are characterized by two stresses with amplitudes $\sigma_{s1}$ and $\sigma_{s2}$; in our simulations, we consider two kinds of stress patterns: (i) a symmetric pattern within the primitive cell, i.e. $\sigma_{s1} = \sigma_{s2}$ and (ii) an asymmetric pattern with $\sigma_{s1} \neq \sigma_{s2}$. Note that our system is characterized by different periodicity scales: the periodicity of graphene $a \simeq 2.46\,\si{\angstrom}$ and the periodicity of the mechanical superlattice, with lattice constant $a_{\sigma} = L \cdot \sqrt{3}$, where $L$ is the distance between the two stress sites inside the primitive supercell (see figure \ref{fig1}). Experimentally accessible situations provide a large size difference between the graphene lattice and mechanical superlattice; in our simulation we consider $L = 3 \,\, {\rm \mu m}$, resulting in a ratio $a/a_{\sigma} \simeq 5 \cdot 10^{-5}$. \\

Simulations are implemented using a commercial FEM solver (COMSOL Multiphysics). The graphene deformations are calculated by a 2D continuum mechanics solver using $E=1\,{\rm TPa}$ and $\nu=0.15$ as the Young modulus and Poisson ratio, respectively~\cite{kudin2001}. Once the strain tensor has been calculated, the pseudo-vector potential $\mathbf{A}_{ps}(x,y)$ and the PMF profile $B_{ps}(x,y)$ are obtained according to equation (\ref{eq:pseudovector}) (see figure~\ref{fig1}). Finally, the electronic states and dispersions are obtained by diagonalizing the 2D Dirac equation within the First Brillouin Zone of the superlattice~\cite{guinea2010,georgi2017}. The eigenvalue problem is solved by imposing $k$-dependent Floquet-Bloch conditions at each side of the superlattice primitive cell~\cite{collet2011}. The periodic condition sets the reciprocal space coordinates within the First Brillouin Zone; the numerical solution of the eigenvalue equation at $(k_x,k_y)$  returns the eigenvalues $E(k_x,k_y)$ from which we build the energy bands. The simulation is performed at few selected stress values, i.e. at selected PMF magnitudes. The strength of the field is expressed in terms of an adimensional parameter $\xi=L/\ell_B$, where $\ell_B = \sqrt{\hbar/eB_{max}}$ is the magnetic length and $B_{max}$ is the maximum $|B_{ps}|$ value. Similarly, energy values will be reported in terms of an adimensional parameter $\tilde{E}$ which is connected to the actual electron energy $E$ by $\tilde{E}=E\sqrt{S}/\hbar v_F$, where $S$ is the surface of the superlattice primitive cell.

\section{Role of angular orientation and identification of flat-band states} \label{sec:symm}

\begin{figure}[]
    \centering
    \includegraphics[scale=0.425]{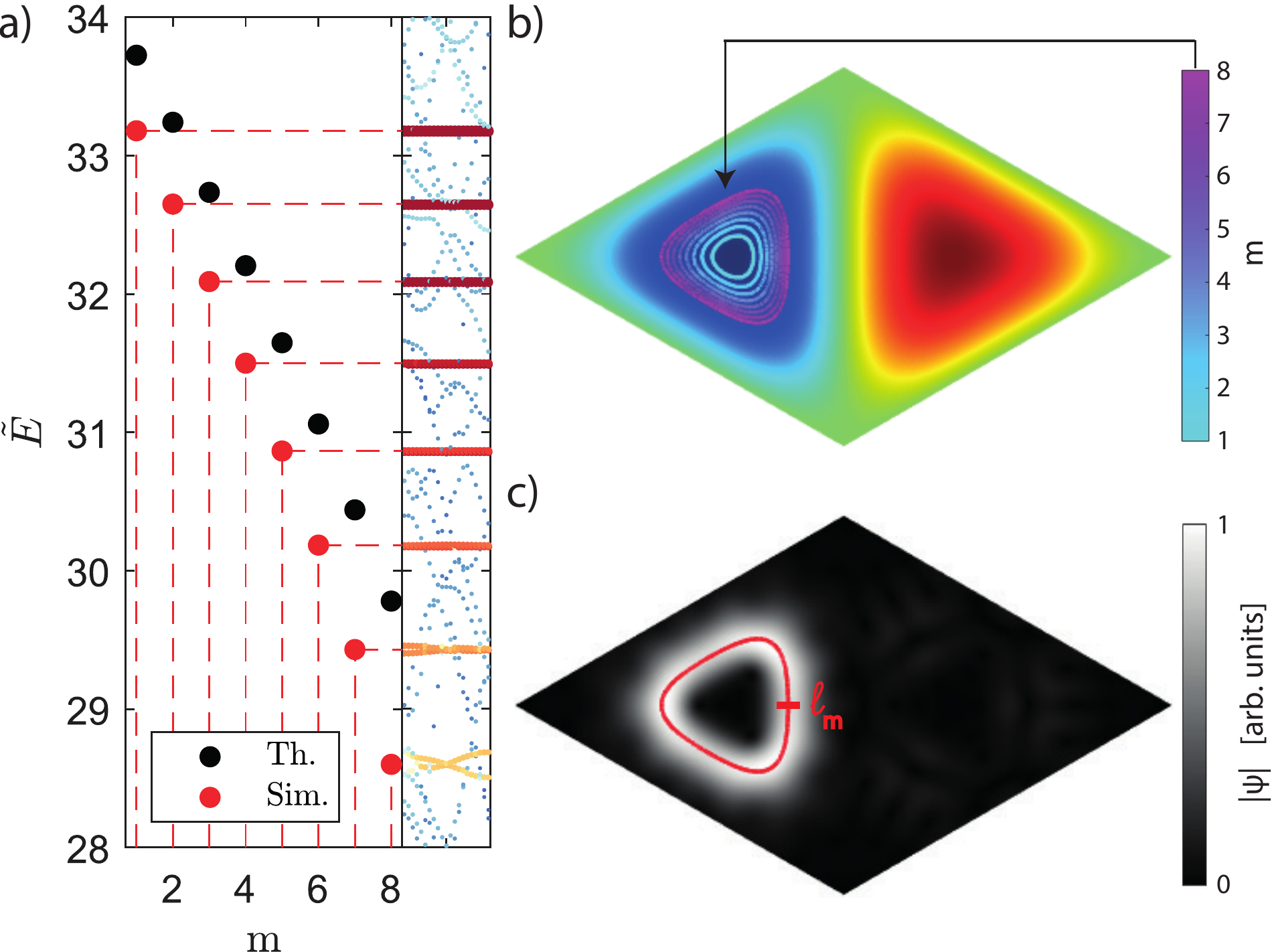}
    \caption{a) Energies of the L1 subbands. Black dots corresponds to results from equation \ref{eq:flux}, while red dots are obtained from the band structure, whose magnified view is reported in the rightmost panel. b) Magnetic orbits (from the condition $B_{ps}=B_m$) for L1 and each sub-band index $m$. The orbits are superposed to the negative PMF (the result for the positive PMF lobe is analogous). c) Magnetic orbit (red line) for n=1 and m=8 superposed to the corresponding wavefunction. The red scale bar corresponds to the magnetic length $\ell_m$, calculated using $B_m$.}
    \label{fig3}
\end{figure}

The geometry and periodicity of the PMF induced by a periodic stress can be modulated through the relative rotation of graphene with respect to the applied stress pattern, which is parametrized by the angle $\theta$ (see figure~\ref{fig1}). As it can be deduced from the three-fold rotational symmetry of equation (\ref{eq:pseudovector}), $2\pi/3$ rotations leave the pseudo-magnetic field unchanged, while $\pi/3$-rotations flip its sign.
The dependence of angular orientation can be summarized considering the two significant angles $\theta = \pi/6$ and $\theta = 0$. The markedly different PMFs resulting from these cases are displayed in figures \ref{fig1}(a) and (b), respectively. In the case $\theta = \pi/6$, the field has two lobes with antinodes located at the stress sites $s1$ and $s2$, as displayed in figure~\ref{fig1}(a). The opposite happens for the case $\theta = 0$ (figure~\ref{fig1}(b)), where the nodes are located at the stress sites s1 and s2. By rotating the superlattice, we can shift the PMF in order to have its nodes strongly/weakly overlapping the stress sites. This could become important in an experimental framework where the disturbance at the stress sites makes the underlying graphene inaccessible or significantly perturbed.\\ 
For each rotation angle, we use solutions of equation (\ref{eq:dirac}) to obtain the density of states (DoS) as a function of the PMF field magnitude. This is displayed in figure~\ref{fig2}(a) for the case $\theta = \pi/6$. As reported in different recent works~\cite{milovanovic2020,moldovan2013}, the evolution for increasing fields is complex, with a variety of states migrating towards lower and higher energies.\\ 
The former ones (see "L0-migrating states" in figure~\ref{fig2}(a)) converge to the zero-energy level, originating from the $n=0$ Landau level which has been observed in graphene in several experiments~\cite{levy2010,lu2012,li2015} using real magnetic fields.\\
The blue-shifting set of states (L1, L2, N0 and S0) follow different trends. A general $\xi$-proportionality can be identified for two sets of states, L1 and L2 (see figure~\ref{fig2}(a)), which forms from around $\xi\sim10$ and nearly follows the trend calculated for $n=1$ and $n=2$ Landau levels in real magnetic field (white dashed lines in figure~\ref{fig2}(a)). This behaviour suggests that L1 and L2 states are related to Landau levels originating from the quasi-uniform PMF, which are called pseudo-Landau levels (pLLs).\\ 
Additionally to the expected pLLs, we identify a low energy peak in the DoS, (N0), which blue-shifts sublinearly with $\xi$ as well as a set of multiple states, (S0), which increase superlinearly with $\xi$.

The yellow-coloured peaks in the DoS of figure~\ref{fig2}(a) correspond to energy bands which are flat in the First Brillouin zone of the supercell (panel (b)). Note that high symmetry points of the superlattice (${\rm \overline{\Gamma}}$, ${\rm \overline{K}}$, ${\rm \overline{K}'}$) have been indicated with the bar notation to not confuse them with the ones used for graphene (${\rm \Gamma}$, ${\rm K}$, ${\rm K'}$). 
To show all the different flat band states we described earlier, we have chosen two significant band structures for $\xi=10,15$ in order to illustrate S0 states (S0$'$ and S0$''$)  and L0, L1, L2, N0, respectively (for further details, see Supplementary Information). The single points in the band structures in figure~\ref{fig2}(b) have been scaled in size and color-coded using two field-confinement parameters: $\eta^-$ and $\eta^+$. These are defined:
\begin{equation}
\eta^\pm = \int_S |\psi(\vec{r})|^2 \cdot |\tilde{B}(\vec{r})| \cdot \Theta[\pm \tilde{B}(\vec{r})] d\vec{r}
\end{equation}
where $\tilde{B}$ is the normalized PMF, $\Theta$ the heaviside function and $S$ is the surface of the superlattice primitive cell. A wavefuction deeply confined within a PMF lobe will have a large $\eta_{tot}=\eta^++\eta^-$. Conversely, a wavefunction localized within the PMF nodes will have a vanishing  $\eta_{tot}$. We observe that, combining $\eta^{\pm}$, the flat states can be operatively classified in terms of their electron confinement, as shown in figure~\ref{fig2}(b).

\begin{figure*}[]
    \centering
    \includegraphics[scale=0.25]{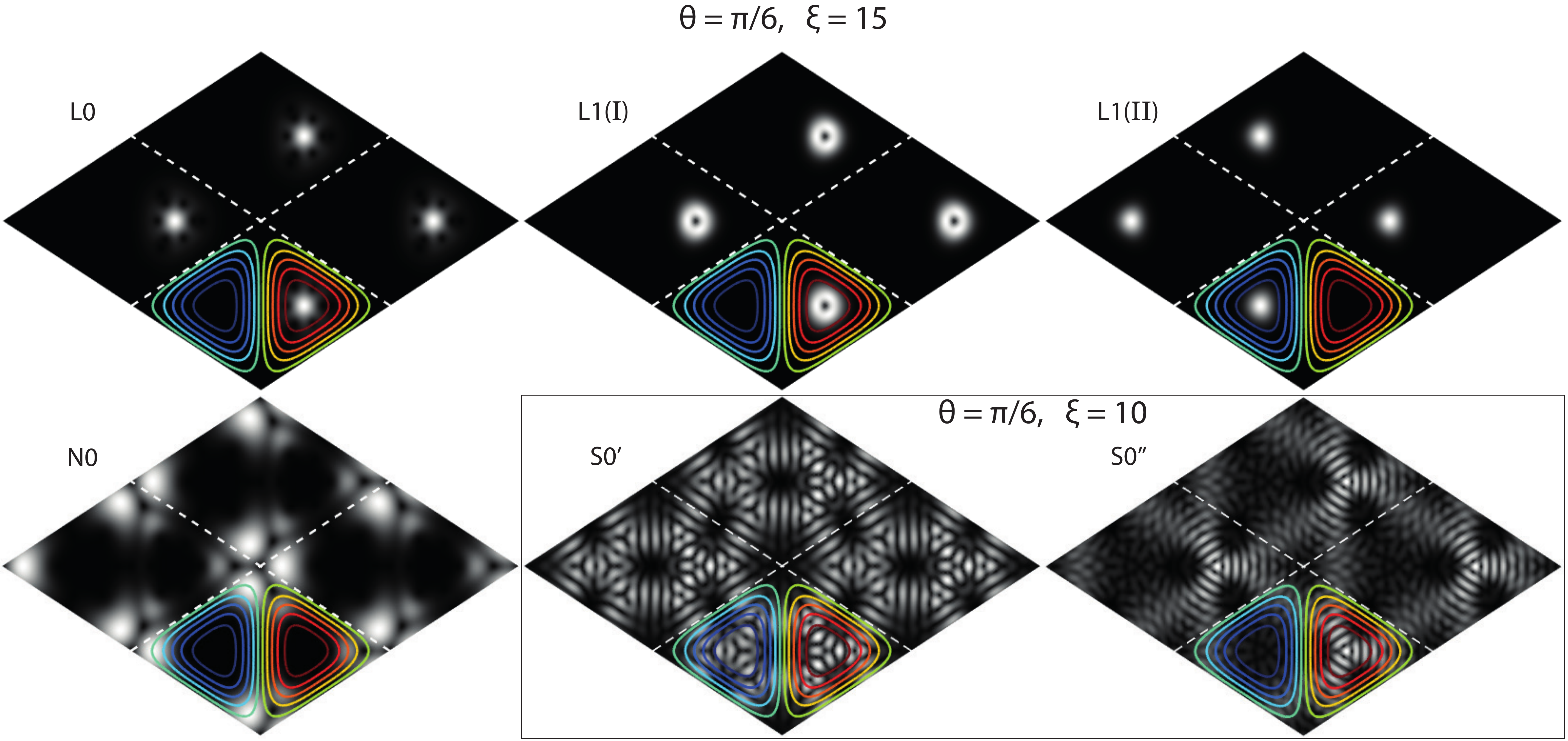}
    \caption{Eigenstates belonging to the set L0, L1, N0 and S0 (identified from green triangles in figure~\ref{fig2}(b)) for the case $\theta = \pi/6$ at the $\overline{\Gamma}$ point. The labels (I) and (II) identify each of the 2 degenerate levels. The contour plot of the pseudo-magnetic field in figure~\ref{fig1}(a) is superimposed to the bottom primitive cell, for each of the eigenstates. Eigenstates L0, L1(I), L1(II) and N0 are taken at $\xi=15$ while eigenstates S0' and S0'' are taken at $\xi=10$ (see green triangles in figure~\ref{fig2}(b)).}
    \label{fig4}
\end{figure*}

We can immediately notice that levels N0 and S0, though being flat and peaked in the density of states, are located outside the field lobes; on the other hand, we observe that the levels L0, L1 and L2 are characterized by peaks in the density of states and high values of $\eta_{tot}$ (high electron confinement inside the field lobes). The confinement confirms their interpretation as pLLs originating from the PMF. As is the case for the Landau levels in graphene in real field, we expect a single degeneracy for level L0; both L1 and L2 present more complex features and by inspecting the energy bands it can be seen that are characterized by a fine structure composed by several sub-bands approximately equispaced in energy. This particular configuration is due to the non-homogeneous nature of PMFs; the spacing of the pLL sub-bands can be estimated by considering the number $m$ of iso-field orbits which enclose an integer magnetic flux ($m\cdot h/e$). The magnetic fields satisfying these relationship, $B_m$, can be found by solving the following equation: 
\begin{equation}
\int_S B_{ps}(\vec{r}) \theta(B_m-B_{ps}(\vec{r})) d\vec{r} - m \frac{h}{e} = 0.
\label{eq:flux}
\end{equation}
The selected $B_m$ can be plugged in the expression for Landau levels in graphene in a uniform and real magnetic field $E_{n,m}=v_F\sqrt{2e\hbar B_m n}$~\cite{neto2009}. The results of this numerical evaluation for the first pLL is displayed in figure~\ref{fig3}(a). The estimated energy spacing is roughly linear and agrees well with the one for the sub-bands, with a small discrepancy within $4\%$. 
A spatial visualization of the first 8 orbits is reported in figure \ref{fig3}(b), superimposed to the PMF profile. The good agreement between direct numerical calculation and the results obtained through equation \ref{eq:flux} can be appreciated in figure~\ref{fig3}(c), where the $m=8$ orbit is superimposed with the corresponding wavefunction. Similar results have been found also for the second pLL, L2.\\
To get some insight on the difference between the pLLs and N0 and S0 states, we plot in figure~\ref{fig4} the corresponding wavefunctions.
The difference in the eigenstates corresponding to the levels S0, N0, L0 and L1 can be clearly seen. As discussed, the pLLs are confined within the PMF regions, with a characteristic wavefunction length-scale given by $\ell_m = \sqrt{\hbar/eB_m}$.  Regarding the L0 level, we expect the electronic wavefunction for a given pseudospin to be localized only within the positive or negative PMF region~\cite{mao2020}. This is exactly what can be seen in figure~\ref{fig4}, where the single degenerate L0 state is confined within positive $B_{ps}$. Changing graphene pseudospin, eigenstate confinement would look exactly the same as in figure~\ref{fig4}, although it would be confined within a negative $B_{ps}$ region. The levels belonging to L1 are characterized by a double degeneracy, i.e. (L1(I), L1(II)). One of the two degenerate levels is confined within the positive field region, while the other is confined in the negative one.\\
In addition to the pLLs, geometrical considerations suggest that electrons can build resonances by multiple field-induced scattering, forming Bloch states which are located outside or just partially overlapping with the PMFs. This is what is observed for states N0 and S0. The former is a flat level completely localized within zero-field regions ($\eta_{tot} \simeq 0.02 $)~\cite{milovanovic2020}. Its peculiar localization is convenient in an experimental framework, being outside the stress sites $s1$ and $s2$ which can strongly perturb the electronic structure of graphene. The states belonging to the S0 set are particularly interesting; their formation can be understood considering the clockwise (counterclockwise) trajectories of electrons in a positive (negative) magnetic field. The electronic path arranged in a infinite periodic system can form closed loops, as are the eigenfunctions shown in figure~\ref{fig4}. Interestingly, these "snake states" have the double property of being spatially delocalized while at the same time having flat bands, implying larger electron correlations due to their low group velocities. While the experimental observation of these states could be hindered by the presence of defects or non-homogeneity of the mechanical superlattice, they could act as linking paths for Landau or pseudo-Landau states which could be arranged into high-coherence networks. Looking at the snake-states at larger energy, the eigenfunctions increase their localization within a PMF of a set sign and therefore have an increasing $\eta^+$ ($\eta^-$) with a vanishing $\eta^-$ ($\eta^+$), forming a network including only one of the two sublattices of the mechanical superlattice. 
A sketch of the different flat-band states we identified is reported in figure~\ref{fig1}(a).\\ 

\begin{figure}[]
    \centering
    \includegraphics[scale=0.45]{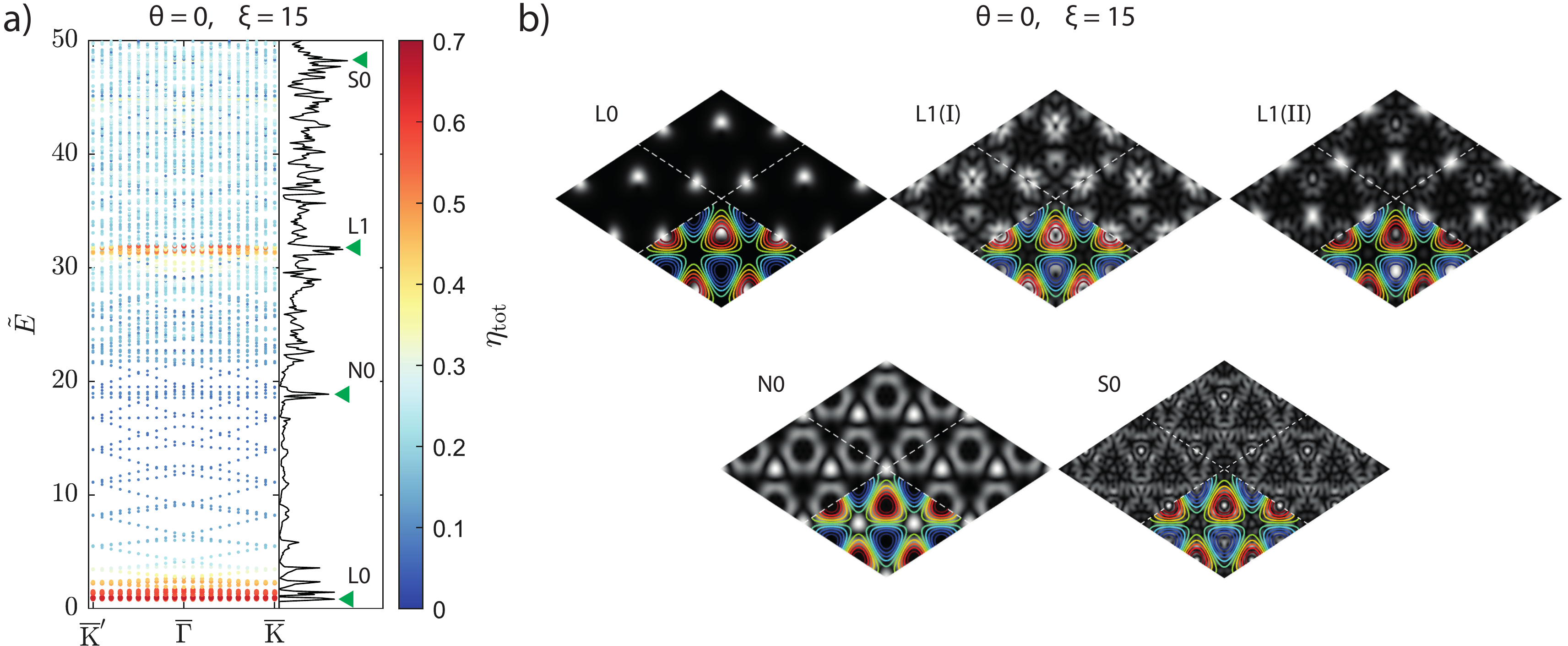}
    \caption{a) Energy band structure and corresponding density of states (cross-section) for $\xi = 15$ in the configuration $\theta=0$. The green triangles in the DoS correspond to the labeled states. b) Eigenstates for the case $\theta=0$ at the $\overline{\Gamma}$ point corresponding to the energy levels highlighted by the green triangles (see panel (a)).}
    \label{fig5}
\end{figure}

In a possible experiment to investigate the states here described, some limitations can arise for states localized in the actuation regions which are mechanically stressed. In most of the local strain engineering techniques~\cite{shioya2014,colangelo2018}, the actuators themselves can strongly perturb the graphene layer or simply make it inaccessible with external out-of-plane probes such as the ones used in STM~\cite{shioya2015} due to electron irradiation of the actuators.
A further tool to address this issue is given by the relative rotation of graphene and the mechanical superlattice. In the analyzed case of $\theta=\pi/6$ rotation, the PMF antinodes coincide with the sites of maximum stress; different rotations could lead to very different situations. In facts, a $\theta=0$ rotation generates the PMF profile shown in figure~\ref{fig1}(b), where the stress sites $s1$ and $s2$ coincide with nodes of the PMF. Moreover, the field lobe extension is reduced with respect to the $\theta=\pi/6$ case (figure~\ref{fig1}(a)). The reduction of a factor $1/\sqrt{3}$ translates into a shift and reduction of the multiplicity of pLLs, as can be seen in the band structure of figure~\ref{fig5}(a). A reduced field extension requires a larger energy in order to form a close orbit and give rise to quantization. 
Significantly, a rearrangement of stress sites can be used to strongly modify the field periodicity and the electronic band structure without modifying the periodicity of the mechanical superlattice. In the $\theta=0$ case, we observe a clear electron confinement of the L0 and L1 pLL (see figure~\ref{fig5}(b)), with features similar to the ones reported for $\theta=\pi/6$. N0 states can also be identified: in this case they have a weak localization within the field region even if most of the wavefunction is confined in the field nodes/underneath the stress sites. The reduced field extension pushes the snake states S0 to lower energies thus, for the case $\theta=0$, they can be found at $\xi=15$ (see an example in figure~\ref{fig5}(b)).

Concluding, we identified and classified several kinds of flat-band states in periodically strained graphene. Along with the well known pLLs, where electrons are localized within the lobes of the field regions,
we identified states strongly localized within the PMF nodes. A further tuning knob for states localization is given by the graphene and stress lattice relative rotation, which switches the overlap of the stress sites with the PMF nodes/antinodes and changes the overall size of field lobes.

\section{Asymmetric stress configuration} \label{sec:asymm}

\begin{figure*}[]
    \centering
    \includegraphics[scale=0.4]{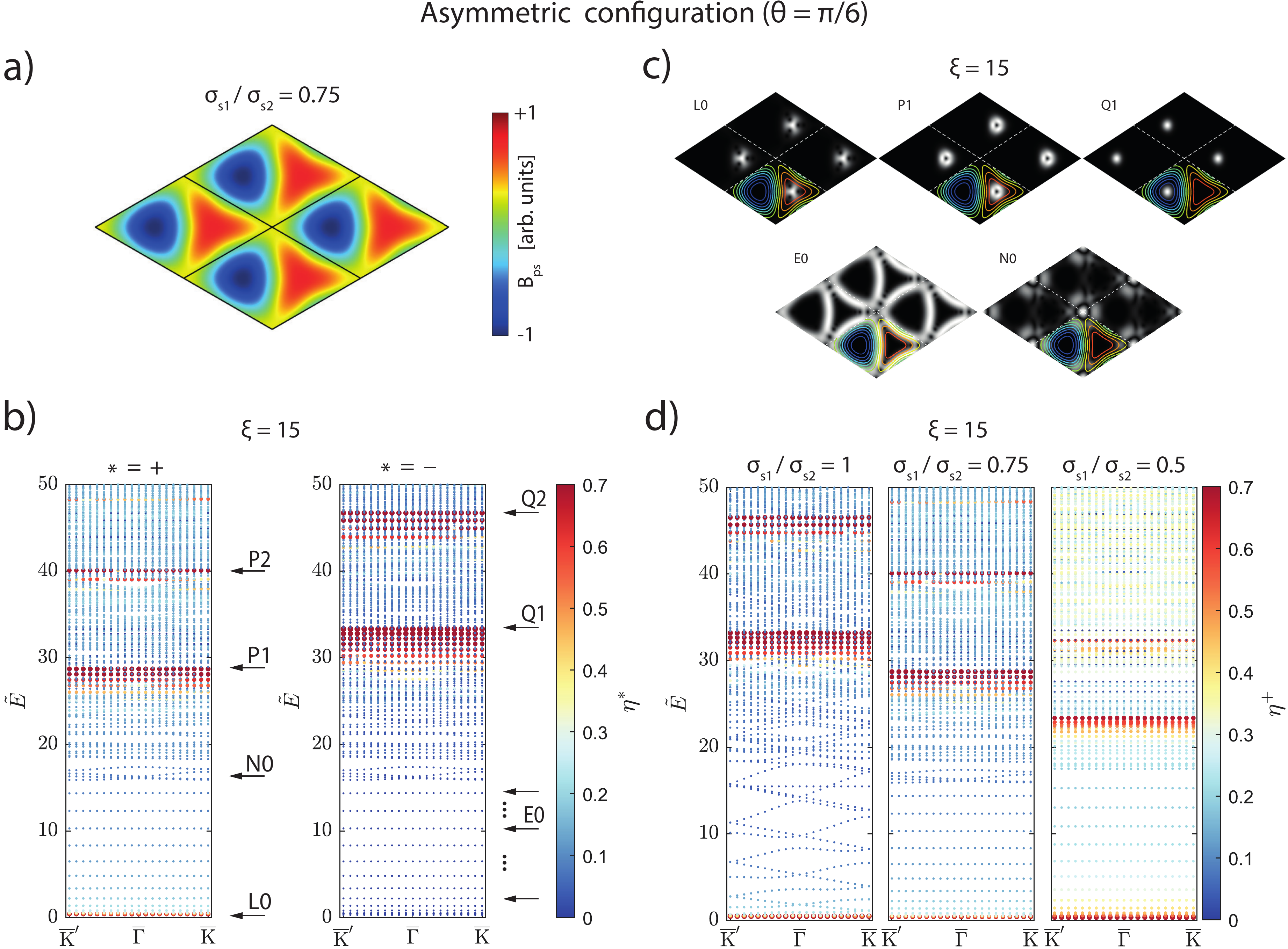}
    \caption{a) Pseudo-magnetic field for $\sigma_{s1}/\sigma_{s2} = 0.75$. b) Electronic band structure ($\xi = 15$) color-coded with $\eta^+$ (left) and $\eta^-$ (right). c) Eigenstates at the $\overline{\Gamma}$ point corresponding to the energy levels highlighted in panel (b). The contour plot of the pseudo-magnetic field in panel (a) is superimposed to the bottom primitive cell, for each eigenstate. d) Evolution of the band structure for decreasing (from left to right) values of the applied stress at the site $s2$.}
    \label{fig6}
\end{figure*}

The possibilities given by infinite strain lattices are not limited by the symmetric honeycomb structure analyzed so far. In the symmetric case, the low-energy electronic bands induced by the superlattice are qualitatively similar to the ones found in graphene layers; our PMF array {\em de facto} defines an artificial graphene material. The most recent strain engineering techniques can go one step further, controlling the single superlattice site in a flexible and reconfigurable way. This allows for the creation of more complex superlattice configurations: by changing the stress of a single site, artificial defects can be defined; by modifying the stress of each site belonging to a sublattice, we can configure~different artifical materials remiscining of structures such as hBN~\cite{Wang2017} which are widely known for having large band gaps. More complex configurations would see a mix of artificial graphene, defected graphene and artificial hBN in planar heterostructures which could lead to advanced electronic transport devices. As a first step in this direction, we show here how the electronic configuration of periodically strained graphene can be modified by increasing the asymmetry of the stress sites belonging to different sublattices.
Let's consider the infinite periodic stress lattice with a rotation of $\pi/6$, as shown in figure~\ref{fig1}(a): an asymmetry can be introduced by considering one of the two stress site as 75\% weaker than the other (i.e. $\sigma_{s1}/\sigma_{s2}=0.75$). The resulting PMF is displayed in figure~\ref{fig6}(a), and the corresponding band structure has been plotted in figure~\ref{fig6}(b). The two panels show the same energy dispersion yet with different color codes: leftmost (rightmost) panel has been colored according to $\eta^+$ ($\eta^-$), highlighting the states confined in the positive (negative) regions of the PMF. We can recognize most of the flat-band states we described so far.

The most striking feature though is the large numbers of a new wide set of states (E0). Energetically localized between L0 and N0, these flat states are separated by full band gaps; the appearance of band gaps after breaking the sublattice symmetry of the periodic superlattice supports the picture of considering the symmetric and asymmetric lattices as artificial graphene and hBN, respectively. Moreover, we argue that tunable band-gaps could have applications for the creation of planar artificial heterostructures by controlling the energy level occupation and electron confinement. 
The states E0 are localized outside the field lobes, forming a ring-like shape network delocalized in the whole crystal, as displayed in figure~\ref{fig6}(c). For the opposite pseudospin the wavefunction is qualitatively similar, with the ring-like shape slightly distorted and disconnected vertex. The E0 states closer to L0 gradually localize inside the positive PMF lobe; on the other hand the ones closer to N0 localize inside the nodes of the PMF (see figure~\ref{fig6}(c)). Regarding the other states, the non-degenerate L0 level is still localized only in the positive PMF, with no equivalent state confined within the negative PMF for the pseudospin we are considering.\\ 
The asymmetry impacts the $n \geq 1$ pLLs. The pseudospin degeneracy is lifted and two sets of pLLs appear, one confined in the positive field (labeled $P$) and the other in the negative one (labeled $Q$). The $P$ and $Q$ states have a net energy difference originating from the different geometries of positive and negative PMFs; in particular, the energy difference $\Delta E= |E_P-E_Q|/E_L$  is about 15\% for both $n=1$ and $n=2$ pLLs. From figure~\ref{fig6}(c), we observe that the wavefunctions of P1 and Q1 are essentially identical to L1(I) and L1(II) shown in figure~\ref{fig5}, albeit with different eigenenergies.

A further weakening of the magnitude of the stress at one site leads to a continuous tuning of the band structure. Figure~\ref{fig6}(d) shows, as an example, a few configurations with increasing degree of asymmetry starting from the symmetric case.  The bands have been color-coded according to $\eta^+$, therefore highlighting $P$ levels. An increasing weakening of the site $s1$ produces a red-shifts of the $P$ states. This behavior is consistent with the fact that the PMF amplitude in positive regions (site $s2$) becomes smaller than the amplitude in negative ones (site $s1$). Conversely, $Q$ states are almost unchanged, being strongly localized in negative field regions whose amplitude is roughly constant for all considered cases. The continuous tuning of the band structure could represent an important tool for tapering planar PMF heterostructures, in such a way to reduce electronic scattering.

\section{Conclusion} \label{sec:conclusion}

In this paper we introduce a method to modulate the electronic states in graphene by employing a superlattice of planar micro-actuators~\cite{shioya2014,colangelo2018}. The stress lattice forms PMF patterns whose periodicity and confinement can be controlled by acting on the relative superlattice-graphene twist. The effect of the PMF array creates several flat-band states which have recently been investigated for a possible exploitation of their increased electron-electron correlations. On top of the well known pLLs, localized within the PMF regions, we identify flat states lying on the field nodes (N0) as well as snake states (S0) generated from resonant conditions of particular geometric configurations. The possibility to overlap or separate the stress sites and the PMF by controlling the graphene-superlattice rotation can be crucial for the experimental observation of these states in regions of pristine graphene, unaffected by the mechanical actuators. 
Finally, we show how breaking the sublattice symmetry opens several band gaps and allows a fine tuning of the pLL flat states, paving the way to the creation of planar heterostructures defined by different stress superlattices.

\section*{Acknowledgments}
This work has been supported in part by the PRIN project MONSTRE-2D of the Italian MIUR.

\bibliographystyle{BSTart}
\bibliography{References}

\newpage
\begin{center}
 {\huge Supplementary Information}
\end{center}
\renewcommand{\thesection}{S\arabic{section}}
\setcounter{section}{0}
\renewcommand{\thefigure}{S\arabic{figure}}
\setcounter{figure}{0}

\section{Mechanical system and Floquet Bloch-conditions}

\begin{figure*}[hbt!]
    \centering
    \includegraphics[scale=0.4]{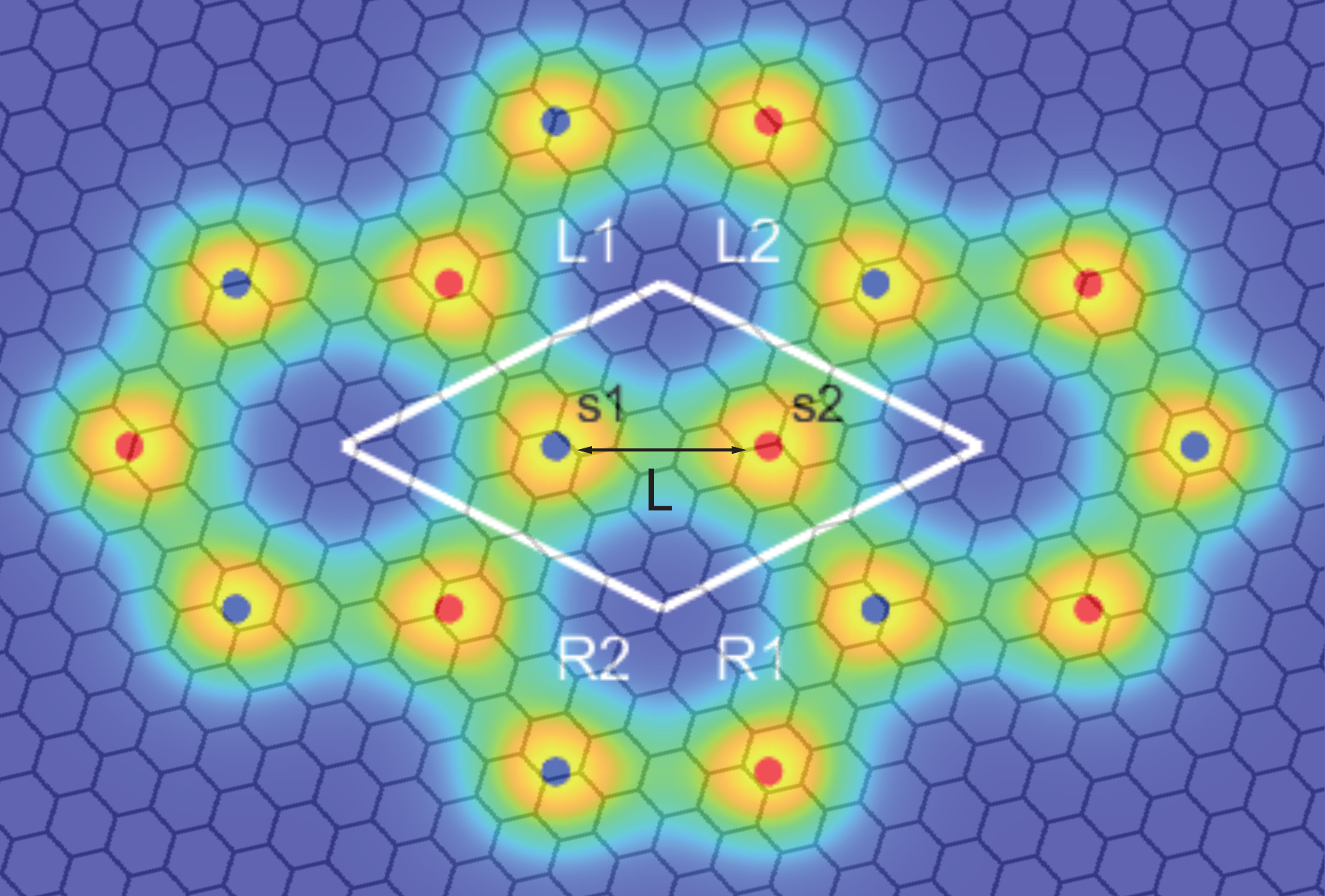}
    \caption{Superlattice stress configuration. b) Strain tensor elements for $\theta=\pi/6$. c) Strain tensor elements for $\theta=0$.}
    \label{fig1_supp}
\end{figure*}

\begin{figure*}[hbt!]
    \centering
    \includegraphics[scale=0.3]{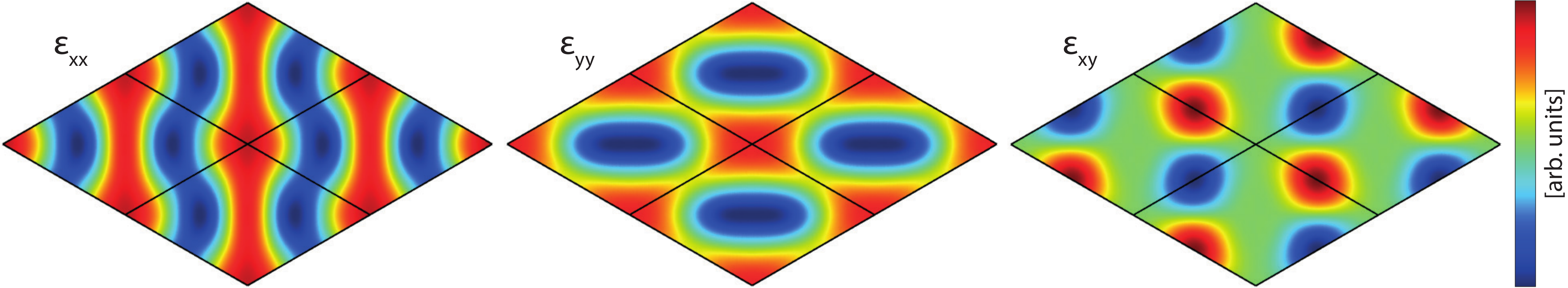}
    \caption{Strain tensor elements.}
    \label{fig2_supp}
\end{figure*}

The stress configuration which is applied in the primitive cell is displayed in figure \ref{fig1_supp}: we apply a set of gaussian inside and outside the primitive in order to regularize the behaviour of the pseudo-magnetic field (PMF) at the edges of the primitive cell. At each stress site $s1$ and $s2$ (identified by blue and red dots, respectively), we apply a set of gaussian and planar stresses with magnitude $\sigma_{s1}$ and $\sigma_{s2}$ respectively. The two sets of magnitude can be independently tuned to control the asymmetry of the stress configuration.
In order to calculate the (PMF) resulting from the periodic stress configuration in figure \ref{fig1_supp}(a), we calculated the elements of the strain tensor $\epsilon_{xx}$, $\epsilon_{yy}$ and $\epsilon_{xy}$ which are displayed in figure \ref{fig2_supp}.\\

To take into account the periodicity of the superlattice, we imposed Floquet-Bloch conditions at each side of the superlattice primitive cell, in particular:
\begin{equation}
\begin{split}
&\psi_{L1} = \psi_{R1} e^{i(k_x x + k_y y)} \\
&\psi_{L2} = \psi_{R2} e^{i(k_x x + k_y y)}
\end{split}
\end{equation}
where each subscript $(L1, R1, L2, R2)$ refers to the edges of the primitive cell of the stress superlattice as reported in figure \ref{fig1_supp}.

\section{Electronic band structure}

\begin{figure*}[hbt!]
    \centering
    \includegraphics[scale=0.4]{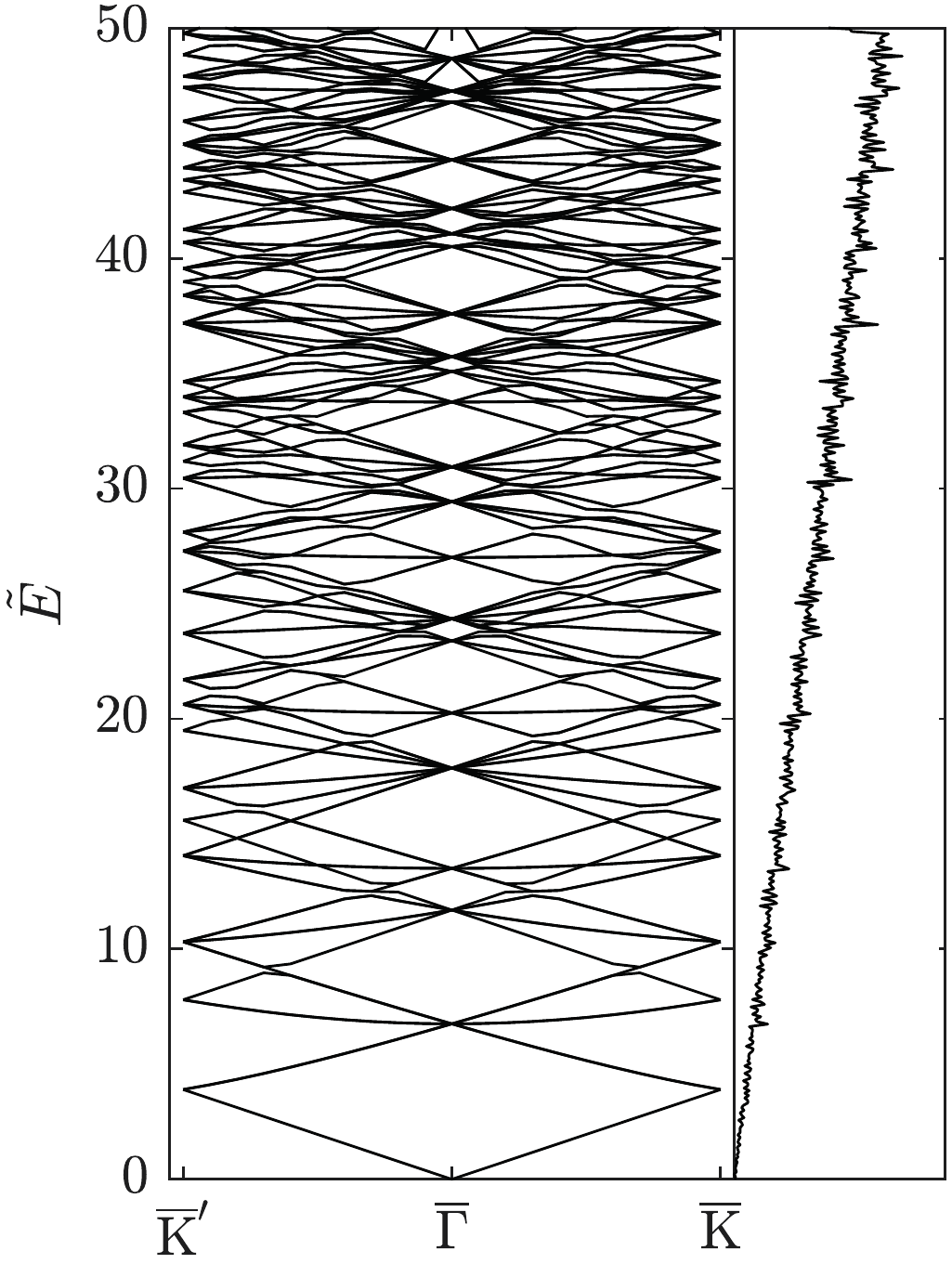}
    \caption{a) Band structure (left) and density of states (right), at $\xi = 0$.}
    \label{fig3_supp}
\end{figure*}

\begin{figure*}[hbt!]
    \centering
    \includegraphics[scale=0.4]{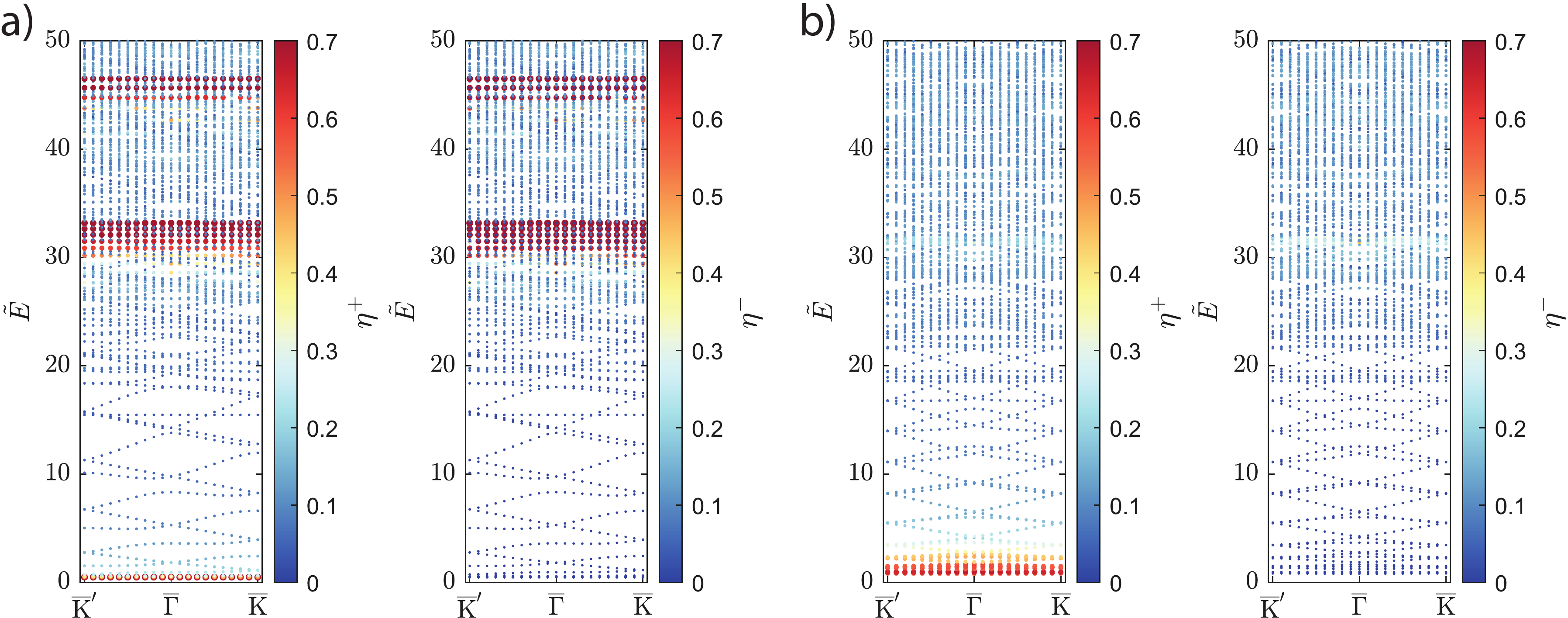}
    \caption{a) Band structure for $\theta = \pi/6$ and $\xi = 15$. b) Band structure for $\theta = 0$ and $\xi = 15$.}
    \label{fig4_supp}
\end{figure*}

In figure \ref{fig3_supp}, we display the band structure (left panel) and the density of states (right panel) calculated at $\xi=0$, i.e. in absence of applied stress.
In figure \ref{fig4_supp}(a) and (b), we display the band structure for the two cases $\theta = \pi/6$ and $\theta = 0$ ($\xi=15$), respectively for the two confinement parameters $\eta^+$ and $\eta^-$.

\section{Mesh}

We observe that the calculated zero Landau level in the band structure is not found exactly at zero energy: in particular, the energy shift of the zeroth Landau level shift increases when the PMF magnitude is increased.
We argue that this effect is due to the finite size of the mesh elements which, at larger PMF magnitudes, becomes comparable with the characteristic length-scale of the PMF, $\ell_B$ (see the main manuscript). In particular, at fixed $\xi$, the energy shift of the zeroth pLL is inversely proportional to the characteristic element size, $\ell_M$, employed during the numerical solution.
The data shown are calculated using $\ell_M/\ell_B \lesssim 0.35$.

\end{document}